\documentclass[pra,twocolumn,showkeys,showpacs,superscriptaddress,floatfix,longbibliography]{revtex4-1}
\usepackage[usenames,dvipsnames]{color}
\usepackage[english]{babel}
\usepackage{graphicx}
\usepackage{color}
\usepackage{amssymb,amsmath}
\usepackage[Gray,squaren]{SIunits}
\usepackage{xspace}
\usepackage{xcolor}
\usepackage{upgreek}
\usepackage{ulem}
\usepackage{epstopdf}
\usepackage{amssymb,amsmath,verbatim,ulem}
\usepackage[export]{adjustbox}
\usepackage{dsfont}
\usepackage{appendix}
\usepackage{comment}
\usepackage[colorlinks = true,
           linkcolor = blue,
           urlcolor  = blue,
           citecolor = blue,
           anchorcolor = blue]{hyperref}
\newcommand{\Eq}[1]{Eq.~(\ref{eq:#1})}
\newcommand{\fig}[1]{Fig.~\ref{#1}}

\newcommand{\ket}[1]{|{#1}\rangle}
\newcommand{\bra}[1]{\langle{#1}|}

\begin{document}

\title{Datta-Das transistor for atomtronic circuits using artificial gauge fields}
\author{Chetan Sriram Madasu}
\affiliation{School of Physical and Mathematical Sciences, Nanyang Technological University, 21 Nanyang Link, Singapore 637371, Singapore.}
\affiliation{MajuLab, International Joint Research Unit IRL 3654, CNRS, Universit\'e C\^ote d'Azur, Sorbonne Universit\'e, National University of Singapore, Nanyang Technological University, Singapore}
\author{Mehedi Hasan}
\altaffiliation[Currently at ]{Cavendish Laboratory, University of Cambridge, Cambridge CB3 0HE, United Kingdom}
\affiliation{School of Physical and Mathematical Sciences, Nanyang Technological University, 21 Nanyang Link, Singapore 637371, Singapore.}
\affiliation{MajuLab, International Joint Research Unit IRL 3654, CNRS, Universit\'e C\^ote d'Azur, Sorbonne Universit\'e, National University of Singapore, Nanyang Technological University, Singapore}
\author{Ketan Damji Rathod}
\altaffiliation[Currently at ]{Bennett University, Greater Noida 201310, India}
\affiliation{Center for Quantum Technologies, National University of Singapore, Singapore 117543, Singapore.}
\author{Chang Chi Kwong}
\affiliation{School of Physical and Mathematical Sciences, Nanyang Technological University, 21 Nanyang Link, Singapore 637371, Singapore.}
\affiliation{MajuLab, International Joint Research Unit IRL 3654, CNRS, Universit\'e C\^ote d'Azur, Sorbonne Universit\'e, National University of Singapore, Nanyang Technological University, Singapore}
\author{David Wilkowski}
\email{david.wilkowski@ntu.edu.sg}
\affiliation{School of Physical and Mathematical Sciences, Nanyang Technological University, 21 Nanyang Link, Singapore 637371, Singapore.}
\affiliation{MajuLab, International Joint Research Unit IRL 3654, CNRS, Universit\'e C\^ote d'Azur, Sorbonne Universit\'e, National University of Singapore, Nanyang Technological University, Singapore}
\affiliation{Center for Quantum Technologies, National University of Singapore, Singapore 117543, Singapore.}

\date{\today}
\begin{abstract}
Spin-dependent electrical injection has found useful applications in storage devices, but fully operational spin-dependent semiconductor electronics remain a challenging task because of weak spin-orbit couplings and/or strong spin relaxations. These limitations are lifted considering atoms instead of electrons or holes as spin carriers. In this emerging field of atomtronics, we demonstrate the equivalent of a Datta-Das transistor using a degenerate Fermi gas of strontium atoms as spin carriers in interaction with a tripod laser-beams scheme. We explore the dependence of spin rotation, and we identify two key control parameters which we interpret as equivalent to the gate-source and drain-source voltages of a field effect transistor. Our finding broadens the spectrum of atomtronics devices for implementation of operational spin-sensitive circuits.
\end{abstract}

\maketitle

\section{Introduction}

Atomtronics encompasses devices with ultracold neutral atoms as the carriers instead of electrons or holes in standard electronics. Unlike electronic devices where only the charge of the carriers bears information in the form of electric current and voltage, atomtronic devices store and manipulate information in internal as well as external states of the atom (see \cite{amico2021roadmap} for a recent review). This capability leads to new devices and sensors based on quantum coherence or entanglement that have no analogues in electronics. Employing state-of-the-art cooling techniques and modern fabrication technologies, it is now possible to produce, guide and manipulate ultracold gasses in dedicated circuits. Examples of such atomtronic devices include: lensing \cite{Sourabh2021MWLensing}, curved guiding beyond adiabatic regime \cite{PhysRevLett.124.250403}, batteries with chemical potential as a source for neutral currents \cite{Caliga2017AtomtronicBattery}, capacitors \cite{lee2013analogs}, switches and quantized supercurrents \cite{Ramanathan2011ToroidalBEC, Wright2013SuperfluidAtomCitcuit,Mossman2019OnewaySwitch}, transistors using triple well potential \cite{Caliga2016TransistorTrippleWell}, and Josephson junctions \cite{luick2020ideal,kwon2020strongly}. Combining these devices, neutral-atom-based quantum networks can be constructed with applications envisioned in quantum technologies such as simulations and sensing \cite{amico2021roadmap}.

In the spirit of the above-mentioned devices, we implemented for atomtronic circuits the equivalent of a Datta-Das transistor (DDT) \cite{Datta1990eEOM}. DDT is the spintronics version of a field effect transistor (FET) with a semiconductor gate region sandwiched between the ferromagnetic source and drain. The working principle of DDT is that spin polarized electrons from the source enter the gate region where their spins precess due to spin-orbit coupling controlled by the gate voltage \cite{Datta2018HowWeProposedSpinTransistor}. The contact between the gate and the drain acts as a spin filter that allows only certain orientation of spins to pass through to the drain, thereby controlling the source to drain spin current. Implementing DDT in solid-state systems proved to be challenging due to multiple difficulties such as; inefficient spin polarization in the source and drain, spin injection into the gate, depolarization of spin and insufficient strength of spin-orbit coupling \cite{Chuang2014AllSemiconductorSpinTransistor}. However, in atomtronic systems, these difficulties do not arise due to well controlled optical pumping, and an almost decoherence-free environment.

\begin{figure}[h!]
    \includegraphics[scale=0.95]{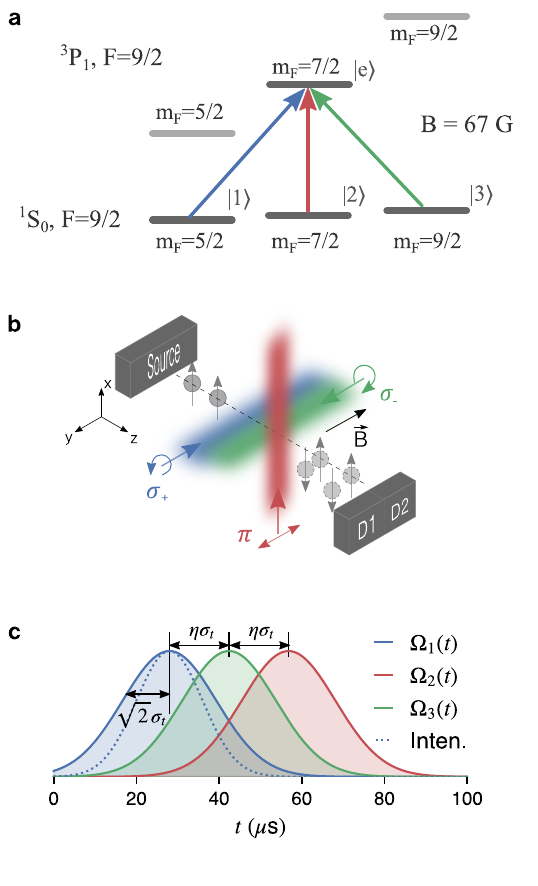}
    \caption{\textbf{Principle of the atomtronics DDT}. \textbf{a} The tripod scheme used in the experiment. A magnetic field of $67\,$G is applied to isolate the Zeeman sub-state $\ket{F_e=9/2,m_F=7/2}$ of the triplet $^3P_1$ excited level. \textbf{b} Schematic showing the operating principle of an atomtronic DDT device with the arrangement of tripod beams as seen from a moving atom at velocity $\boldsymbol{v}=v_z \hat{\boldsymbol{e}}_z$. Spin polarized atoms emanating from an atomtronic source (left black box) enter the spin-orbit coupled gate region and their spin is rotated. Depending on their final spin they arrive at a different location in the drain (right black box). The polarization of each the beam is indicated in the figure where we use the following notation convention $1\equiv\sigma_+$, $2\equiv\pi$, and $3\equiv\sigma_-$. \textbf{c} Rabi frequencies of the tripod lasers as a function of time in the atom's proper frame with $\sigma_t=8\ \mu$s and $\eta=1.8$. The normalized laser beam intensity of $\sigma_+$ (standard deviation of $\sigma_t$) is shown as a blue-dotted curve.}
    \label{Fig1}
\end{figure}

Our atomtronic version of a DDT is implemented using a resonant tripod scheme on a degenerate Fermi gas. Spin-polarized atoms enter an interacting region consisting of three Gaussian beams and exit with their spin orientation controlled on-demand through the peak power of one beam \cite{Vaishnav2008DDT}. We show that the spin rotation is insensitive to a wide range of initial atomic velocity distribution, making the system relevant as an atomtronic device. Moreover, because of a transverse spatial separation of the two spin states at the DDT output, a spin filter is naturally built into the system.

\section{Implementation} 

We prepare an ultracold gas of 87-strontium ($^{87}\textrm{Sr}$) atoms using a two stage magneto-optical trap (MOT) cycle, cooling first using the $^1S_0\to$ $^{1}P_1$ transition at $461\,$nm (frequency linewidth $32\,$MHz) followed by the $^1S_0 \to$ $^3 P_1$ intercombination line at $689\,$nm (frequency linewidth $7.5\,$kHz), see Refs. \cite{chaneliere2008three,Tao2015HighFlux} for details. After the laser cooling stage, $N\sim 2.5 \times 10^6$ atoms are transferred into a crossed beams optical dipole trap. Atoms in positive $m_F$ states of the ground state are optically pumped to the $m_F=9/2$ stretched state whereas the negative $m_F$ states are left untouched. Forced evaporative cooling is performed on the partially polarized gas for $5.5\,$s by lowering the powers of the dipole trap beams \cite{hasan2022anisotropic}. We obtain a degenerate Fermi gas with $T/T_F = 0.25$ ($T_F$ is the Fermi temperature associated with the $m_F=9/2$ population) with a temperature of $T\sim 50\,$nK and $N_{9/2}\sim 4\times10^4$ atoms. Moreover, the gas is in a sub-recoil temperature regime, meaning that $\sigma_v<v_r$, where $v_r=\hbar k/m\sim 6.5\,\textrm{mm/s}$ is the single-photon recoil velocity and $\sigma_v\sim\sqrt{k_BT_F/m}\sim 4.3\,\textrm{mm/s}$ is the standard deviation of the Fermi degenerated gas velocity distribution. $\hbar$, $k$, and $m$ are the reduced Planck constant, the wavenumber of the tripod beams, and the atomic mass, respectively.  After evaporative cooling, the optical dipole trap is switched off and a magnetic field bias of $67\,$G is turned on to isolate a tripod scheme on the $^1S_0, F_g=9/2 \to$ $^3 P_1, F_e=9/2$ hyperfine transition of the intercombination line, as shown in Fig. \ref{Fig1}a. Three Gaussian beams resonantly couple the three internal ground states $|a \rangle \equiv |F_g, m_F\rangle$, with $a=1,2,3$ and $m_F=5/2, 7/2, 9/2$, respectively, to a common excited state $\ket{e}\equiv\ket{F_e, m_F=7/2}$. The light-atom coupling is characterized by three Rabi frequencies amplitude $\Omega_a$ associated to the $\ket{a}\to\ket{e}$ transition, respectively \cite{Frederic2018Non-Abelian}.

The operating principle of our atomtronics DDT device is depicted in \fig{Fig1}b. An atom, prepared in the state $\ket{3}$, departs towards the interacting region and moves at constant velocity $\boldsymbol{v}=v_z \hat{\boldsymbol{e}}_z$. The atom crosses the tripod beams with Gaussian shape in the following tripod beams order: 1 ($\sigma_+$), 3 ($\sigma_-$), and 2 ($\pi$). The beams propagate in different direction in the $xy$-plane with polarizations that satisfy the electric dipole transition selection rules. The beam propagation axes are equally spaced with along the $z$-axis. In the proper inertial frame of the atoms, which in the experiment coincides with the laboratory frame, the Rabi-frequency amplitude of the tripod beams have the form of Gaussian pulses parameterized as
\begin{align}
    \Omega_i(t) = \Omega_{0i}e^{-(t-t_i)^2/4\sigma_t^2},
\end{align}
with $t=r/v_z$ and $r$ the atomic center-of-mass position. $\Omega_{0i}$ is the value if peak Rabi frequency, and $\sqrt{2}\sigma_t$ is its temporal standard deviation. We define $\eta$ as the pulse separation in units of $\sigma_t$, so $t_1=3.5\sigma_t$, $t_2=t_1+2\eta\sigma_t$ and $t_3=t_1+\eta\sigma_t$ correspond to the three mean times, where the time origin is taken when the optical dipole trap is switched off, see \fig{Fig1}c. 

\begin{figure}[h]
    \includegraphics[scale=0.78]{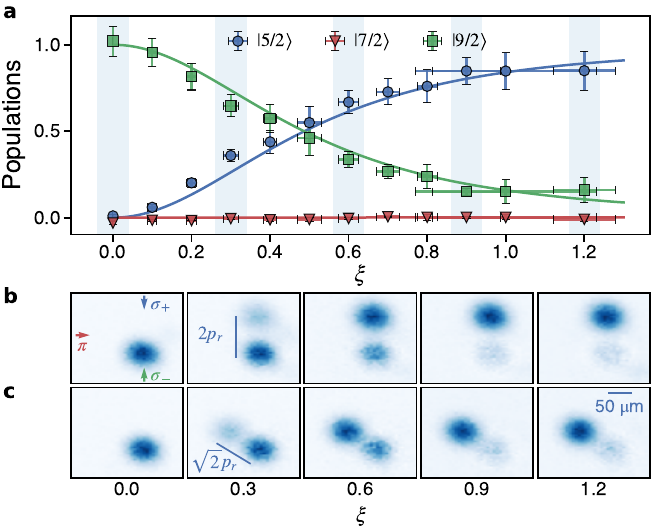}
    \caption{\textbf{Atomtronics DTT output populations. a} Ground-state populations at the output of the interacting region as a function of the gate parameter, $\xi=\Omega_{03}/\Omega_{01}$ for laser pulse sequence shown in Fig. \ref{Fig1}c. Here $\Omega_{01}=\Omega_{02}\approx2\pi\times 270\,$kHz. The plain curves correspond to the model prediction using the unitary operator of \Eq{unitaryT}. \textbf{b} Fluorescence images of the cloud  after the laser pulses and a time of flight of $9\,$ms. The images are associated with the shaded data points on panel a). \textbf{c} Fluorescence images of the cloud  when the order of pulse sequence is reversed i.e., in the order $2$, $3$ and $1$.}
    \label{Fig2}
\end{figure}

\section{Control of the pseudo-spin rotation} 

The populations of the tripod ground states after a Gaussian pulse sequence with $\sigma_t=8\ \mu$s and $\eta=1.8$ as a function of $\xi=\Omega_{03}/\Omega_{01}$ ($\Omega_{02}=\Omega_{01}$) are shown in \fig{Fig2}a. They are extracted from  fluorescence images taken after $9\,$ms of time of flight \cite{hasan2022anisotropic}, see \fig{Fig2}b. During the coupling with the tripod beams, modifications of the ground state populations are mediated by coherent Raman transitions involving a change of the atoms net momentum as the beams have different propagation directions \cite{Kasevich1991AIWithRamanTransitions, Bataille2020SU(N)Imaging, Aspect1989VelocityCPT,wilkowski2009low}. Therefore, an atom initially in the state $\ket{3}$ is coherently transferred to the state $\ket{1}$ by absorbing one photon from beam $3$ and undergoing stimulated emission of one photon into beam $1$. During this process the atom acquires a net momentum of $2p_r\hat{\boldsymbol{e}}_y$, where $p_r=mv_r$ is the recoil momentum. Hence, the distribution with a mean momentum $2p_r\hat{\boldsymbol{e}}_y$ is identified to be in the state $\ket{1}$, whereas the untouched distribution at zero mean momentum is associated to the state $\ket{3}$. Similarly, an atom could absorb one photon from beam $3$ and undergo stimulated emission of one photon into the beam $2$ and being transferred to the state $\ket{2}$ with a net momentum $p_r(\hat{\boldsymbol{e}}_y-\hat{\boldsymbol{e}}_x)$. In the configuration corresponding to \fig{Fig1}c with results depicted in \fig{Fig2}a, the state $\ket{2}$ remains unpopulated at the DDT output. However, by swapping the temporal order of beams $1$ and $2$, we reverse the role played by the states $\ket{1}$ and $\ket{2}$ and found population at a net momentum $p_r(\hat{\boldsymbol{e}}_y-\hat{\boldsymbol{e}}_x)$ corresponding to the state $\ket{2}$, whereas here the state $\ket{1}$ remains unpopulated at the DDT output. This situation is depicted in \fig{Fig2}c, and shows the symmetric role played by the states $\ket{1}$ and $\ket{2}$. Therefore, two output momentum ports are available by simply swapping beams $1$ and $2$, as it was previously pointed out with a similar beam configuration in Ref. \cite{Theur1999MWBS}. One notes that this swapping between states $\ket{1}$ and $\ket{2}$ also occurs by changing the propagation direction of the atoms. The momentum-spin dependence offers a natural solution for filtering the final pseudo-spin current since the pseudo-spin components are spatially separated after ballistic propagation. On the contrary, if pseudo-spin states are required to be in the same momentum state, one could for example prevent momentum exchanges by reversing the direction of beam $1$ to make it co-propagates with beam $3$.

To understand qualitatively our atomtronic DDT device, we identify three important features. First, the temporal variation of the pulse needs to be slow enough such that the dynamics remains adiabatic, consequently the excited state $\ket{e}$ is not populated at any time. Hence, no spontaneous emission occurs and the evolution remains coherent and time-independent. Second, if the final pulse is $\Omega_2$, it means that the output state does not contain the state $\ket{2}$, like for any stimulated Raman adiabatic passage (STIRAP) experiments \cite{Vitanov2017STIRAPReview}. Therefore, the state $\ket{2}$ can be viewed as an ancilla, and the states $\ket{1}$ and $\ket{3}$ form the pseudo-spin basis of the DDT. Finally, the second pulse $\Omega_3$ is coupled to the input state $\ket{3}$ and its amplitude acts as a pseudo-spin-orientation control knob. Indeed, if the pulse amplitude is zero, the input state will not be coupled and the pseudo-spin retains its initial orientation. On the other limiting case where $\Omega_{03}$ is large, the state should be transferred to $\ket{1}$, corresponding to a $\pi$-rotation of the pseudo-spin. In the intermediate range, the experiment depicted in \fig{Fig2}a indicates non-zero populations in both states $\ket{1}$ and $\ket{3}$. We test that evolution is coherent by applying, immediately after the first pulse sequence, an second pulse sequence in the reversed order, namely 2, 3, and 1, and observe that more than $85\%$ of the population returns back to the initial input state $\ket{3}$. Moreover, non-adiabatic processes are likely populating the excited state $\ket{e}$ and lead to incoherent spontaneous emission which will change the net momentum of atoms in a random manner. This decoherence mechanism can be traced in the experiment by monitoring the total population of the two output distributions with defined net momentum shown in \fig{Fig2}b$\&$c with respect to the input $\ket{3}$ population. If the ratio $\rho$ = output population in peaks/input population decays significantly, it means that incoherent processes are at play. For the cases shown in \fig{Fig2}, $\rho = 0.90(2)$, indicating that incoherent processes are unlikely.  

\section{Theory model and approximations} 
\label{Sec_TM}

For quantitative comparison, we describe now a model inspired by a former theoretical work \cite{Vaishnav2008DDT}. We consider the operating principle of our atomtronics DDT device as depicted in \fig{Fig1}b. The Hamiltonian of the tripod scheme in the interaction representation reads $H_I=\frac{1}{2}\sum_{a=1}^3\tilde{\Omega}_a(\textbf{r})\ket{e}\bra{a}+\textrm{H.c.}$, with $\tilde{\Omega}_1(\textbf{r})=\Omega_1(\textbf{r})\textrm{e}^{-iky}$, $\tilde{\Omega}_2(\textbf{r})=\Omega_2(\textbf{r})\textrm{e}^{ikx}$, and $\tilde{\Omega}_3(\textbf{r})=\Omega_3(\textbf{r})\textrm{e}^{iky}$. After diagonalization of $H_I$, we find two zero-eigenenergy degenerated dark states and two bright states with eigenenergies light shifted by $\pm\hbar\Omega/2$, with $\Omega=\sqrt{\sum^3_{a=1}\Omega_a^2}$ \cite{Unanyan1998STIRAP}. The dark states in the bare states basis read
\begin{eqnarray}
\label{eq:D}
\ket{D_1}&=&\sin{\beta}\textrm{e}^{2iky}\ket{1}-\cos{\beta}\textrm{e}^{ik(y-x)}\ket{2}\\
\ket{D_2}&=&\cos{\alpha}(\cos{\beta}\textrm{e}^{2iky}\ket{1}+\sin{\beta}\textrm{e}^{ik(y-x)}\ket{2})-\sin{\alpha}\ket{3},\nonumber
\end{eqnarray}
where  $\alpha=\arctan{(\sqrt{\Omega_1^2+\Omega_2^2}/\Omega_3)}$ and $\beta=\arctan{(\Omega_2/\Omega_1)}$. Importantly, we note that at the input and output of the device, the dark states are connected to bare states as $\lim_{z \to \pm\infty} \ket{D_2} =\lim_{\alpha \to \pi/2}\ket{D_2} = \ket{3}$ and
$\lim_{z \to +\infty} \ket{D_1} =\lim_{\beta \to \pi/2}\ket{D_1} = \ket{1}$.
We now consider that the system evolves in the subspace spanned by degenerate dark states and the bright states are adiabatically eliminated. The state vector reads $\ket{\Psi}=\sum_{j=1,2}\psi_j\ket{D_j}$, where the spinor wavefunction $\underline{\Psi}=(\psi_1,\psi_2)^{\intercal}$ obeys the Schr\"odinger equation, \cite{Ruseckas2005Non-Abelian,Dalibard2011Colloquium}
\begin{equation}
i\hbar\frac{\partial\underline{\Psi}(\mathbf{r},t)}{\partial t}=\left[\frac{(-i\hbar\boldsymbol{\nabla}-\mathbf{\hat{A}}(\mathbf{r}))^2}{2m}+W(\mathbf{r})\right]\underline{\Psi}(\mathbf{r},t).
\label{eq:M_Scho_S}
\end{equation}
The vector and scalar gauge field potentials ($\mathbf{\hat{A}}, \hat{W}$) are represented by $2\times 2$ matrices with  entries $\mathbf{\hat{A}}_{jk}=i\hbar\langle D_j\vert\boldsymbol{\nabla}D_k\rangle$ and $\hat{W}_{jk}=\left[\hbar^2\langle \boldsymbol{\nabla}D_j\vert \boldsymbol{\nabla}D_k\rangle-(\mathbf{\hat{A}}^2)_{jk}\right]/2m$ ($j,k = 1,2$), see Sec. \ref{Sec_MM}. To address the problem in proper frame of the atom, we perform a Galilean transformation to a frame moving at $\boldsymbol{v}$. The general form of the Schr\"odinger equation is preserved using the transformations \cite{brown1999galilean,Vaishnav2008DDT,hasan2022anisotropic}: $\textbf{r}_p =\textbf{r}-\boldsymbol{v}t,\, t_p =t,\quad \boldsymbol{\nabla}_p =\boldsymbol{\nabla},\,\frac{\partial}{\partial t_p} =\frac{\partial}{\partial t}+\boldsymbol{v}\cdot\boldsymbol{\nabla},\, \mathbf{\hat{A}}_p =\mathbf{\hat{A}},\, \hat{W}_p =\hat{W}-\boldsymbol{v}\cdot\mathbf{\hat{A}}$, whereas the spinor picks up a phase factor as $
\underline{\Psi}_p=\textrm{e}^{i\left(\boldsymbol{v}^2t/2-\boldsymbol{v}\cdot\mathbf{r}\right)m/\hbar}\underline{\Psi}.$
Here the subscript $p$ refers to the proper frame. 

We consider the limiting case where $|\boldsymbol{v}|\gg v_r$, and $\eta\sigma_t\ll (k\sigma_v)^{-1},(kv_r)^{-1} $. The first inequality implies that the atomic wave-packet velocity is well defined and does not significantly change during the interaction with the tripod beams. The second inequality implies, within the interaction time, that the atoms do not significantly explore the phases of the tripod beams due to transverse motion. This could be a stringent condition because it imposes an upper bound on the interaction time which in turn should be compatible with the adiabatic limit. As we will see below, this condition usually, but not always, holds in our experiment. If both inequalities are fulfilled, and the temporal coherence of the atomic wave-packet is much shorter than the temporal extension of a tripod beam pulse, we have $\langle \boldsymbol{v}\cdot\mathbf{\hat{A}}\rangle\gg \langle (i\hbar\boldsymbol{\nabla}-\mathbf{\hat{A}})^2\rangle/2m, \langle W\rangle$, where $\langle \cdots \rangle$ stands for the expectation value (see Sec. \ref{Sec_MM}). The gauge field reduces to its $z$-component (along $\boldsymbol{v}$) $\hat{A}_z(t)=\hbar\cos{\alpha(t)\partial\beta(t)/(v_z\partial t)}\hat{\sigma}_y$, and the evolution operator for the spinor reads 
\begin{equation}
\hat{U}(t_f)=\exp{\left[-i\int_0^{t_f}\cos{\alpha(t)}\frac{\partial\beta(t)}{\partial t}\hat{\sigma}_y\textrm{d}t\right]},
\label{eq:unitaryT}
\end{equation}
where $\hat{\sigma}_y$ is the Pauli matrix along $y$ and $t_f$ is the interaction duration. Using \Eq{unitaryT} and \Eq{D}, we extract the bare state populations as represented by the plain curves in \fig{Fig2}a. 

\begin{figure}[h]
    \includegraphics[scale=0.78]{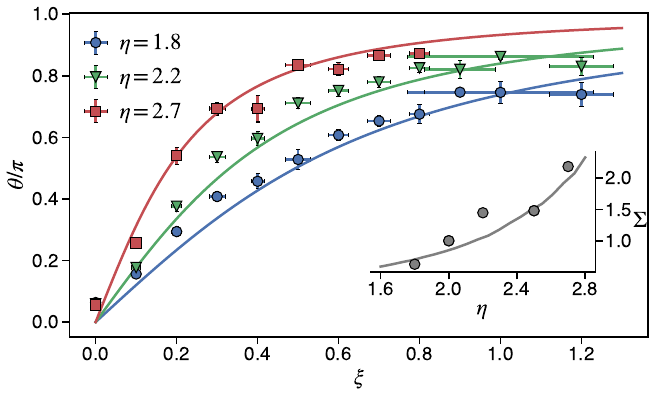}
    \caption{\textbf{Sensitivity of the pseudo-spin rotation.} The polar angle $\theta$ as function of $\xi$ for three different values of the pulse separation $\eta$ for $\sigma_t = 8\ \mu s$. The other experimental parameters are the same as \fig{Fig2}. Inset: $\Sigma = \Delta \theta/\Delta \xi|_{\theta=\pi/2}$ indicates the sensitivity of the pseudo-spin rotation as function of $\eta$. The solid curves are the theory prediction using the unitary operator of \Eq{unitaryT}.}
    \label{Fig3}
\end{figure}

\section{Characterization of the atomtronics DDT} 

From \Eq{unitaryT}, we notice that the pseudo-spin rotation occurs along the $y$-axis, and can be characterized by the polar angle $\theta$ on the Bloch sphere where each pole corresponds to one of the dark state defined by \Eq{D}. By definition
\begin{equation}
\frac{\theta}{2}=\arctan\left(\frac{|\langle D_1| \underline{\Psi}\rangle|}{|\langle D_2| \underline{\Psi}\rangle|}\right)\simeq\arctan\left(\frac{|\langle 1| \underline{\Psi}\rangle|}{|\langle 3| \underline{\Psi}\rangle|}\right),
\label{eq:theta}
\end{equation}
for a measurement after the tripod interaction. In \fig{Fig3}, this polar angle is plotted as function of $\xi$ for three values of the pulse separation $\eta$. We observe a more sensitive rotation when $\eta$ increases. For a more precise analysis of this phenomena, we defined the sensitivity of the pseudo-spin rotation with respect to $\eta$ as the slope $\Sigma(\eta) = \Delta \theta/\Delta \xi|_{\theta=\pi/2}$. The inset in \fig{Fig3} shows that the sensitivity of the pseudo-spin rotation increases monotonically with $\eta$. 

Comparing now our tripod system to a FET, we interpret the pulse separation $\eta$ as the of gate-source voltage of the FET, which increases the slope of the drain-source current over the drain-source voltage in the FET linear region \cite{millman1967electronic}. In this line of thought, the parameter $\xi$ appears to be equivalent to the drain-source voltage.
For practical atomtronics DDT operations, the $\xi$ value shall be the main control parameter, and one could choose a configuration with a large $\eta$ value for spin modulation purposes and with a low $\eta$ value for a precise control of the spin orientation.  

\begin{figure}[h]
    \includegraphics[scale=0.76]{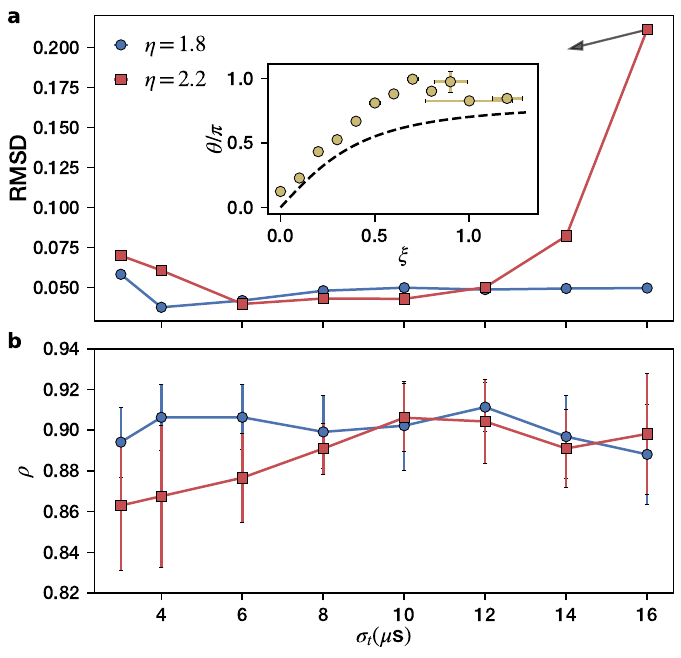}
    \caption{\textbf{Compressing or inflating the gate region. a} RMS deviation of the experimental data from the theoretical prediction using the unitary operator of \Eq{unitaryT}, is plotted as a function of $\sigma_t$ for two values of $\eta$ are plotted. Large RMSD values indicate that the model approximations are not sufficient to capture the experimental behaviour. Inset shows such an example for $\eta=2.2$, and $\sigma_t= 16\ \mu s$. The dashed curve correspond to the prediction. \textbf{b} $\rho=$ output population/input population as a function of $\sigma_t$ for the two values of $\eta$ indicated in panel a).} 
    \label{Fig4}
\end{figure}

As pointed out in an earlier work with similar beams configuration \cite{Theur1999MWBS}, the atomtronics DDT device can also be viewed as a matter-wave beam splitter. This beam splitter is a key element for matter-wave interferometry either in Ramsey-Bord\'e configurations for optical frequency measurement \cite{oates2000improved} or in Mach-Zehnder configurations for inertial sensing \cite{Kasevich1991AIWithRamanTransitions}. In the vast majority of matter-wave interferometers, the beam splitter functionality is ensured by Raman or Bragg transitions \cite{geiger2020high}, which require a precise control of the interaction time, since the pseudo-spin rotation results from Rabi flopping, proportional to the pulse area. In contrast, our matter-wave beam splitter is based on a geometric transformation in the parameter space, and it should be independent of interaction time. Considering the operating principle of our atomtronic DDT device (see \fig{Fig1}b), it means also that the device is expected to be insensitive to the atomic velocity. Because of the unavoidable velocity dispersion of the atomic gas, this property is of fundamental interest. We illustrate this point in \fig{Fig4}a, where we plot the root-mean-squared deviation (RMSD $ = \sqrt{\langle(\theta_{\text{exp}} - \theta_{\text{theory}})^2\rangle}/ \pi$) of the experimental data to the theoretical prediction using \Eq{unitaryT} as a function of $\sigma_t$ for two values of $\eta$. The control parameter $\sigma_t$, which compresses or inflates the interaction zone, is understood to be inversely proportional to the velocity of the atoms. To directly compare the experimental data and the theory prediction, we recall that \fig{Fig3} correspond to $\sigma_t=8\,\mu$s. The RMSD is minimum and close to zero for a range covering a factor 2 to 3 in value of $\sigma_t$ showing that atoms with a broad range of velocities can be targeted by the DDT. For the $\eta=2.2$ case, RMSD increases for both larger and smaller values of $\sigma_t$. This is due to the failure of the model approximations in these two limiting cases. For short $\sigma_t$ value, we note that the value of $\rho$ decreases [see \fig{Fig4}b], which indicates that some spontaneous emission events have taken place, and the adiabatic approximation might no longer be valid. For large $\sigma_t$ values, the adiabatic condition should hold but we observe that the experimental data clearly deviate from the model prediction (see inset). In this case, the inequality $\eta\sigma_t\ll (k\sigma_v)^{-1}$ does not hold, and the velocity spread of the atoms in $xy$-plane induces extra vector potential components that are sensitive to relative phase variations of the tripod lasers, should also be considered \cite{Jacob2007DyninNon-AbelianGF, Ruseckas2005Non-Abelian}. From \fig{Fig4}a, we note that a smaller value of $\eta$ helps to broaden the model validity range in the large $\sigma_t$ limit.

\section{Discussion}

We simulated an atomtronics DDT using a spin-polarized ultracold gas coupled to a sequence of three optical pulse beams addressing a tripod excitation scheme. Thanks to the geometrical nature of the interaction, we showed that the DDT functionalities are independent of the atom velocity for a broad range of values covering the usual velocity dispersion of an ultracold gas. To characterize the behavior of the DDT, we explored two key control parameters of the system, namely the central pulse amplitude and the pulse separation. The coherent rotation of the pseudo-spin occurs due to the presence of an inhomogeneous Abelian gauge field. In addition, photons transfer in the tripod beams cause momentum-spin dependence and conduct to a spatial separation of the pseudo-spin components like in Stern-Gerlach experiment, but of different nature than a spin-Hall effect, which depends on the atoms velocity \cite{Beeler2013SpinHall}. 

Our configuration is suitable for precise spin amplitude control or modulation purpose. However, other arrangements of the tripod beams lead to other kind of output states, which in a general version provides an universal single qubit quantum gate operation \cite{Toyoda2013HolonomicGate,Frederic2018Non-Abelian}. More specifically, instead of having two beams counter-propagating but co-propagating, the momentum-spin dependence is removed and the DDT device controls only the polar angle of the pseudo-spin. The azimuthal angle of the pseudo-spin can also be addressed if one considers two tripod beams overlapping as reported in Ref. \cite{Frank2003Superposition}. Finally we stress that the experiment was performed with a degenerate Fermi gas of strontium atoms, but can be generalized to other type of fermions or bosons as far as a resonant tripod scheme can be isolated, as for example in Rb atoms \cite{PhysRevA.77.023824}.

\section{Materials and Methods: Model derivation}
\label{Sec_MM}

Our starting point is the Schr\"odinger equation given by \Eq{M_Scho_S} governing the spinor wave function $\underline{\Psi}=(\psi_1,\psi_2)^{\intercal}$ evolution in the dark-state manifold. The wave vector is $\ket{\Psi}=\sum_{j=1,2}\psi_j\ket{D_j}$, where $\ket{D_j}$ are the dark states given by \Eq{D}. The spinor dynamics is due to  vector and scalar gauge field potentials represented by matrices with entries $\mathbf{\hat{A}}_{jk}=i\hbar\langle D_j\vert\boldsymbol{\nabla}D_k\rangle$ and $\hat{W}_{jk}=\left[\hbar^2\langle \boldsymbol{\nabla}D_j\vert \boldsymbol{\nabla}D_k\rangle-(\mathbf{\hat{A}}^2)_{jk}\right]/2m$ ($j,k = 1,2$). The vector potential reads
\begin{eqnarray}
    \label{eq:GA}
    \boldsymbol{A}_{11} &=& p_r\left[\cos^2\!\beta\,\hat{\boldsymbol{e}}_x-(1+\sin^2\!\beta)\hat{\boldsymbol{e}}_y\right]\nonumber\\
    \boldsymbol{A}_{12} &=&\boldsymbol{A}_{21}^*=-\cos\alpha\left[\frac{p_r}{2}\sin (2\beta)(\hat{\boldsymbol{e}}_x+\hat{\boldsymbol{e}}_y)+i\hbar \frac{\partial\beta}{\partial z}\hat{\boldsymbol{e}}_z\right]\nonumber\\
    \boldsymbol{A}_{22} &=&  p_r\cos^2\!\alpha\left[\sin^2\!\beta\,\hat{\boldsymbol{e}}_x-(1+\cos^2\!\beta)\hat{\boldsymbol{e}}_y\right].
\end{eqnarray}
The expressions of $\hat{W}$ and $\mathbf{\hat{A}}^2/m$ can be found in \cite{Frederic2018Non-Abelian, hasan2022anisotropic}. We note that some matrices components are proportional to $v_rp_r$ and other to $m^{-1}(\hbar\partial\beta/\partial z)^2$. As indicated in Sec. \ref{Sec_TM}, we perform a Galilean transformation moving at $\boldsymbol{v}$, the center-of-mass velocity of the gas. The Hamiltonian takes the form:
\begin{eqnarray}
    \label{eq:GH}
    H= \frac{(\mathbf{\hat{p}}-\mathbf{\hat{A}})^2}{2m}+W-\boldsymbol{v}\cdot\mathbf{\hat{A}}.
\end{eqnarray}
We consider that $\boldsymbol{v}\gg\sigma_v,v_r$, and we now derive the conditions such that the last right-hand side term of the Hamiltonian \Eq{GH} is dominant. First, we recall that $\boldsymbol{v}=v_z \hat{\boldsymbol{e}}_z$, so $\boldsymbol{v}\cdot\mathbf{\hat{A}}=v_z\hat{A}_z$. Hence, according to \Eq{GA}, $\langle \boldsymbol{v}\cdot\mathbf{\hat{A}}\rangle\sim\hbar v_z\partial\beta/\partial z=\hbar\partial\beta/\partial t\simeq\hbar(\eta\sigma_t)^{-1}$, where $\eta\sigma_t$ is the temporal extension of a tripod beam pulse. Since, $\langle\mathbf{\hat{p}}^2\rangle\sim (\sigma_v/m)^2$, and $\sigma_v<v_r$, we find immediately that $\langle \boldsymbol{v}\cdot\mathbf{\hat{A}}\rangle\gg\left[\mathbf{\hat{p}}^2/(2m),\,(\mathbf{\hat{p}}\cdot\mathbf{\hat{A}})_{x,y}/m\right]$ if $\eta\sigma_t\ll(kv_r)^{-1}$. The same inequality holds if one considers the $\hat{W}$ and $\mathbf{\hat{A}}^2/m$ terms proportional to $v_rp_r$. We now find that the inequality $\langle \boldsymbol{v}\cdot\mathbf{\hat{A}}\rangle\gg(\mathbf{\hat{p}}\cdot\mathbf{\hat{A}})_{z}/m$ is fulfilled if $\boldsymbol{v}\gg\sigma_v$, which is our primary hypothesis. Finally, we can check that $\boldsymbol{v}\cdot\mathbf{\hat{A}}$ is dominant over the $\hat{W}$ and $\mathbf{\hat{A}}^2/m$ terms proportional to $m^{-1}(\hbar\partial\beta/\partial z)^2$, if $\eta\sigma_t\gg\hbar(mv_z^2)^{-1}$. This inequality means that the temporal coherence of the atomic wave-packet is much shorter than the temporal extension of a tripod beam pulse. Similarly, considering the DDT device as depicted in \fig{Fig1}b, the inequality means also that the atomic wavepacket is well-localized in the the tripod beam pulse. 

If the hypothesis, discussed above, is fulfilled, the Hamiltonian on \Eq{GH} takes the simple form of a spin-orbit coupling term:
\begin{eqnarray}
    \label{eq:GH2}
    H= -v_z\hat{A}_z,
\end{eqnarray}
and we recover the unitary evolution operator of \Eq{unitaryT}.

In our system we have $(k\sigma_v)^{-1}\simeq 50\,\mu$s and  $(kv_r)^{-1}\simeq 16\,\mu$s. Therefore the model discussed above shall be valid if $\eta\sigma_t\ll 16\,\mu$s. Furthermore, the adiabatic condition imposes $\langle \boldsymbol{v}\cdot\mathbf{\hat{A}}\rangle\ll\hbar\Omega$, where $\hbar\Omega$ is the characteristic dark/bright states energy splitting. The inequality is recast as $\eta\sigma_t\gg\Omega^{-1}\simeq 1\,\mu$s. Importantly, the model validity conditions and the adiabatic condition could be fulfilled simultaneously.


%

\end{document}